\documentclass[12pt]{iopart}
\usepackage{times}
\usepackage{bm}
\usepackage{epsfig}
\usepackage{graphicx}
\usepackage{graphics}
\usepackage{amssymb}
\usepackage{color}

\begin{document}
\title[Double Ionization of Atoms by XUV Laser Pulses]
      {Time-Dependent B-Spline R-Matrix Approach to Double Ionization of Atoms by XUV Laser Pulses}
\author{Xiaoxu Guan$^{\dag,1}$, O Zatsarinny$^{\dag,2}$, C J Noble$^{\dag,\S,3}$, K Bartschat$^{\dag,4}$, and \\
        B I Schneider$^{\ddag,5}$}
\address{$^{\dag}$Department of Physics and Astronomy, Drake University, Des Moines, IA 50311, USA}
\address{$^\S$Computational Science and Engineering Department, Daresbury Laboratory, Warrington WA4 4AD, UK}
\address{$^\ddag$Physics Division, National Science Foundation, Arlington, Virgina 22230, USA}

\eads{$^1$xiaoxu.guan@drake.edu, $^2$oleg.zatsarinny@drake.edu, $^3$cjn@maxnet.co.nz, $^4$klaus.bartschat@drake.edu, 
        $^5$bschneid@nsf.gov}
\submitto{\jpb}

\begin{abstract}
We present an {\it ab initio\/} and non-perturbative time-dependent approach to the problem of
double ionization of a general atom driven by intense XUV laser pulses. After using a highly
flexible $B$-Spline $R$-matrix method to generate field-free Hamiltonian and electric dipole matrices,
the initial state is propagated in time using an efficient Arnoldi-Lanczos scheme.
Test calculations for double ionization of He by a single laser pulse yield good agreement
with benchmark results obtained with other methods. The method is then applied to two-color
pump-probe processes, for which momentum and energy 
distributions of the two outgoing electrons are presented.
\end{abstract}


\section{Introduction}
The two-photon double ionization (DI) of the helium atom induced by intense
short XUV laser pulses has received considerable attention from both theorists 
and experimentalists alike.  Instead of listing a large number of references here, we 
note that much of the recent work 
was quoted in recent papers \cite{Feist2008,Guan2008a}. Even within the past
few months, however, several additional papers appeared.

Given the intensities and lengths of the laser pulses involved,
the numerical approaches used to tackle this problem are all essentially attempts to
solve the time-dependent Schr\"odinger equation (TDSE), beginning with
a well-defined initial state before the laser strikes and then propagating
this state in the presence of the laser field by one of a number of numerical approaches.
Once the laser is switched off, various probabilities and, in some cases, generalized
cross sections can be extracted.

Over the past two years, our group has been working on the development of a general
{\it ab initio\/} theoretical approach, which is applicable to complex targets 
beyond (quasi) two-electron systems.  In two recent papers~\cite{Guan2007,Guan2008b} 
we outlined how field-free Hamiltonian and electric dipole matrices generated with the highly flexible
$B$-Spline $R$-matrix (BSR)~\cite{Zat06} suite of codes may be combined
with an efficient  Arnoldi-Lanczos time propagation scheme to describe the interaction of short intense laser pulses
with a complex atom, leading to multi-photon excitation and {\it single\/} ionization.
The key points of our method are the following: 1)~We employ the BSR code,
which allows for the use of non-orthogonal orbital sets, to generate field-free Hamiltonian
and dipole matrices.  2) We then set up an efficient Arnoldi-Lanczos scheme to propagate the 
initial state in time. 3) Finally, we extract the information by standard projection schemes.

In the present paper, we report on the extension of this approach and the corresponding computer code~\cite{CPC08} 
to allow for two electrons in the
continuum and hence the possibility to describe double ionization processes. After outlining the general
method, we present a test application to the He problem, for which many benchmark results
are available for comparison.  Finally, we use the method to investigate pump-probe processes involving
two XUV laser pulses whose characteristics, including a time delay, are assumed to be controlled separately
in the corresponding experiment.

Unless specified otherwise, atomic units (a.u.) are used throughout this manuscript.

\section{THEORY}

\subsection{The TDBSR Approach}

The ingredients of an appropriate theoretical and computational formulation require
an accurate and efficient generation of the Hamiltonian and the electron$-$field interaction matrix elements, as well as
an optimal approach to propagate the TDSE in real time.
As mentioned above, there have been numerous calculations for two-electron systems such as He and also H$_2$.
While these investigations emphasize the important role of two-electron systems in studying electron$-$electron
correlation in the presence of a strong laser field, in its presumably purest form, experiments
with He atoms are difficult and other noble gases, such as Ne and Ar, are often favored
by the experimental community.

Fully {\it ab initio\/} theoretical approaches, which are applicable to complex targets beyond (quasi) two-electron systems,
are still rare.  For (infinitely) long interaction times, the $R$-matrix Floquet ansatz~\cite{Floquet} has been
highly successful. A critical ingredient of this method is the general atomic \hbox{$R$-matrix}
method developed over many years by Burke and collaborators in Belfast. A modification of the method, allowing
for relatively long though finite-length pulses was described by Plummer and Noble~\cite{Plum03}. 
Recently, a time-dependent formulation~\cite{BurBur} of this method was applied to short-pulse laser interactions with
Ar~\cite{vandH2007} and a pump-probe XUV scenario for
Ne~\cite{Lysaght2008}.

Following our recent work on excitation and single-ionization of Ne~\cite{Guan2007} and Ar~\cite{Guan2008b}, 
we now describe what needs to be done to extend our approach to {\it double\/} ionization.
We start with the time-dependent Schr\"odinger equation
\begin{equation}
\label{TDSE}
i \frac{\partial}{\partial t} \Psi(\bm{r}_1,...,\bm{r}_N;t) =
  \big[ H_0(\bm{r}_1,...,\bm{r}_N) + V(\bm{r}_1,...,\bm{r}_N;t) \big] \Psi(\bm{r}_1,...,\bm{r}_N;t)
\end{equation}
for the $N$-electron wavefunction $\Psi(\bm{r}_1,\ldots,\bm{r}_N;t)$,
where $H_0(\bm{r}_1,...,\bm{r}_N)$ is the field-free Hamiltonian containing the kinetic energy of the $N$
electrons, their potential energy in the field of the nucleus, and
their mutual Coulomb repulsion, while
\begin{equation}
\label{length}
V(\bm{r}_1,\ldots,\bm{r}_N;t) = \sum_{i=1}^N \bm{E}(t) \cdot  \bm{r}_i
\end{equation}
represents the interaction of the electrons with the laser field $\bm{E}(t)$ in the dipole length form.
\par\smallskip
The tasks to be carried out in order to
computationally solve this equation and to extract the physical information of interest are:
\newcounter{c3}
\begin{list}{\arabic{c3}.}{\usecounter{c3}}
\item Generate a representation of the field-free Hamiltonian and its eigenstates; 
      these include the initial bound state, other bound states, auto\-ionizing states, as well as
      single-continuum and double-continuum states to represent electron scattering from the residual ion.
\item Generate the dipole matrices to represent the coupling to the laser field.
\item Propagate the initial bound state until some time after the laser field is turned off.
\item Extract the physically relevant information from the final state.
\end{list}

Of particular interest in the experiments mentioned above are processes, in which one, two, or even more electrons
undergo significant changes in their quantum state in the presence of an atomic core. 
These include excitation, single and double ionization,
ionization plus simultaneous excitation, or inner-shell ionization with subsequent rearrangement in the hollow ion.
The latter processes, in particular,
can only be investigated in systems beyond the frequently studied two-electron helium atom or H$_2$ molecule.
This generalization to two electrons
outside a multi-electron core is far from trivial, but the flexibility of the BSR method is highly advantageous
for the tasks~1 and~2.
In contrast to many other approaches~\cite{Hu-Collins-1, Hu-Collins-2, KTaylor-et-al},
our method is formulated in a sufficiently general way to be applicable to {\it complex\/} atoms, such as inert gases
other than helium and even open-shell systems with non-vanishing spin and orbital angular momenta.  
In reality, of course, the size of the problems that can be handled is also determined by the available
computational resources.

The solution of the TDSE requires an accurate and efficient generation
of the Hamiltonian and electron$-$field interaction matrix elements. In order to achieve this goal,
we approximate the time-dependent wavefunction as
\begin{equation}
\Psi(\bm{r}_1,...,\bm{r}_N;t) \approx \sum_q C_q(t) \Phi_q(\bm{r}_1,...,\bm{r}_N).
\label{expansion}
\end{equation}
The $\Phi_q(\bm{r}_1,...,\bm{r}_N)$ are a
set of time-independent $N$-electron states formed from appropriately symmetrized products of
atomic orbitals. They are expanded as
\begin{equation}
\fl
\Phi_q(\bm{r}_1,...,\bm{r}_N) = {\cal A}  \sum_{c,i,j} a_{ijcq} \Theta_c(x_1,\ldots,x_{N-2};
      \hat{\bm r}_{N-1}\sigma_{N-1};\hat{\bm r}_{N}\sigma_{N}) R_i(r_{N-1})R_j(r_{N}).
\label{channel}
\end{equation}
Here ${\cal A}$ is the anti-symmetrization operator, the
$\Theta_c(x_1,\ldots,x_{N-2}; \hat{\bm r}_{N-1}\sigma_{N-1};\hat{\bm r}_{N}\sigma_{N})$
are channel functions involving the space and spin coordinates ($x_i$) of $N\!-\!2$ core electrons
coupled to the angular ($\hat{\bm r}$) and spin ($\sigma$) coordinates of the two outer electrons,
$R_i(r)$ is a radial basis function, and the $a_{ijcq}$ are expansion coefficients.
Although resembling a close-coupling ansatz with two
continuum electrons, the expansion~(\ref{channel}) contains bound states and singly ionized
states as well.  In general, the atomic oritals, $R_i(r)$, are not orthogonal to one another or 
to the orbitals used to describe the atomic core.
If orthogonality constraints are imposed on these functions, additional terms would need to be added 
to the expansion to relax the constraints.  This possibility still exists as an option in our computer code.

In addition to simplifying the expansion, a significant advantage of not forcing such orthogonality
conditions is the flexibility gained by being able to tailor the
optimization procedures to the individual neutral, ionic, and continuum orbitals.
In the BSR code, the outer orbitals (i.e., the $R$~functions above) are expanded in $B$-splines. Factors
that depend on angular and spin momenta are separated from the radial degrees of freedom through the construction
of the channel functions.  Since
many Hamiltonian matrix elements share common features, this enables the production of a
``formula tape'', resulting in an
efficient procedure to generate the required matrix elements.

When the expansion~(\ref{expansion}) is inserted into the Schr\"odinger equation, we obtain
\begin{equation}
i \bm{S} \frac{\partial}{\partial t} \bm{C}(t) = \big[\bm{H}_0 + E(t) \bm{D} \big] \cdot  \bm{C}(t),
\label{non-ortho-tdse}
\end{equation}
where $\bm{S}$ is the overlap matrix of the basis functions, $\bm{H}_0$ and $\bm{D}$ are matrix representations of
the field-free Hamiltonian and the dipole coupling matrices, 
and $\bm{C}(t)$ is the time-dependent coefficient vector in~Eq.~(\ref{expansion}).

The price to pay for the flexibility in the BSR approach, at least initially, is the representation 
of the field-free Hamiltonian and the dipole matrices in a non-orthogonal basis.
In our previous work~\cite{Guan2007}, we described two methods of combining the BSR method with
a highly efficient Arnoldi-Lanczos propagation scheme.  In the first one, we diagonalized the
overlap matrix~$\bm{S}$ and transformed the problem back to an orthogonal basis before applying the
standard propagation scheme~\cite{Park-Light,Schneider-Collins}.  Alternatively, the propagation
can be done more directly in the non-orthogonal basis.  This only requires a Cholesky decomposition of~$\bm{S}$,
but some additional operations at every time step.  

A third approach involves a transformation into the eigenbasis by solving the field-free generalized eigenvalue problem first. 
Details can be found in Guan \etal\cite{Guan2008b}. This method, also used by Laulan and Bachau~\cite{Laulan2003}, has the 
major advantage that the transformation makes it possible to cut unphysically high eigenvalues and the corresponding eigenvectors
from the time propagation scheme, thereby making the scheme more stable. At the same time, it simplifies the definition of the initial 
state and the extraction of the physically interesting information, and it once again allows for the use of the standard Arnoldi-Lanczos 
time propagation scheme. It was applied in the present work as well.  We checked that the results were stable against a variation
of both the cut-off energy (5.5\,a.u. for the results shown below) and the
time-step (100 steps per optical cycle).

We first tested our method for the He problem, i.e., two electrons in the presence of a bare He$^{2+}$ core, and will show some example results below. 
With a box size of 60~a.u. and 81 $B$-splines distributed over this range, with a spacing of 1.0 a.u.~near the edge of the box and
smaller steps near the origin, the ranks of the field-free Hamiltonian blocks ranged between 10,000 and 20,000.  We expect these
ranks to increase to 50,000$-$100,000 for a complex target like Ne. Fortunately, it has become rather  
straightforward to diagonalize matrices of that size and perform the transformation of the dipole matrices to the 
eigenbasis on massively parallel computing platforms.  

\subsection{Extraction of the cross sections in two-photon double ionization process}
Details regarding the extraction of the  
fully differential (energies and angles resolved) as well as the total cross section 
were given by Guan \etal\cite{Guan2008a} and therefore will not be repeated here.
To make this paper self-consistent, however, we summarize the most important formulas used for 
single-color processes and then indicate what has to be done in the case of more than one laser.

In a time-dependent formulation, the dependence of the $N$-photon cross
section on the number of photons absorbed not only occurs through the factor 
$(\omega/I_0)^N$, where $\omega$ is the angular frequency and $I_0$ is the peak intensity of the laser,
but also through the effective interaction time.  
The result for the latter depends explicitly on the number of photons being absorbed. For two
photons it is given by~\cite{Foumouo2006}
\begin{equation}
T_{\rm eff}^{(N)}\equiv \int_0^{\tau} f^{2 N}(t) \, {\rm d}t
\end{equation}
with the result
\begin{equation}
T_{\rm eff}^{(2)} = 35\tau/128 
\end{equation}
for a sine-squared pulse shape of duration of $\tau$.  

The energy sharing between the two escaping electrons can be uniquely determined
through the hyperangle \hbox{$\alpha \equiv \tan^{-1}(k_2/k_1)$}. We set
\hbox{$E_1 = E_{\rm exc}\cos^2\alpha$} and \hbox{$E_2 = E_{\rm exc}\sin^2\alpha$}, where 
\hbox{$E_{\rm exc}=2\omega- I^{2+}$} is the excess energy for two-photon double ionization.
For a given energy sharing $\alpha$, the triple-differential cross section (TDCS) can be
written as
\begin{equation}
\fl
\frac{{\rm d}^{3}\sigma}{{\rm d}\alpha{\rm d}\hat{\boldsymbol{k}}_1{\rm d}
\hat{\boldsymbol{k}}_2} =
\bigg(\frac{\omega}{I_0}\bigg)^2
\frac{1}{T_{\mathrm{eff}}^{(2)}}\int{\rm d}
k'_1{\rm d} k'_2 
k_1'^2k_2'^2 
\,\delta(\alpha-\tan^{-1}\big(\frac{k'_2}{k'_1})\big)
\big|\langle \Psi_{\boldsymbol{k}'_1, \boldsymbol{k}'_2}^{(-)}|\Psi(t)\rangle
\big|^2,
\end{equation}
where $|\Psi(t)\rangle$ is the time-propagated wave\-function.  
Also, the singlet two-electron continuum wave\-function satisfying the incoming-wave boundary
condition ($-$) is given by
\begin{equation}
\Psi_{\boldsymbol{k}_1, \boldsymbol{k}_2}^{(-)}
(\mbox{\boldmath $r$}_1,\mbox{\boldmath $r$}_2)=
\frac{1}{\sqrt{2}}\bigg[
\Phi^{(-)}_{\boldsymbol{k}_1 }(\boldsymbol{r}_1 )
\Phi^{(-)}_{\boldsymbol{k}_2 }(\boldsymbol{r}_2 )
+
~\Phi^{(-)}_{\boldsymbol{k}_2 }(\boldsymbol{r}_1 )
\Phi^{(-)}_{\boldsymbol{k}_1 }(\boldsymbol{r}_2 )
\bigg].
 \label{eq:continuum}
\end{equation} 
With the one-electron Coulomb function given by
\begin{equation}
\Phi^{(-)}_{ \boldsymbol{k} }(\mbox{\boldmath $r$})
=\frac{1}{k}\sum_{lm}{\mbox
i}^{l}e^{-{\rm i}\sigma_{l}(k)}\varphi^{(c)}_{kl}
(r)Y_ { lm }
(\hat{\mbox{\boldmath $r$}} )Y_{lm}^{*}
(\hat{\mbox{\boldmath$k$}}),
\label{momentum-partial}
\end{equation}
and the asymptotic behavior
\begin{equation}
\varphi^{(c)}_{kl}(r) ~~\longrightarrow~~ \hskip-1.1truecm\lower3.0truemm\hbox{\footnotesize $r 
\to \infty$} ~\sqrt{\frac{2}{\pi}}\sin\big(kr+\frac{Z}{k} \ln2kr-\frac {l\pi } { 2 } +\sigma_l\big),
\end{equation}
with $\sigma_l$ as the Coulomb phase,
the two-electron Coulomb function~(\ref{eq:continuum}) is normalized in
momentum space according to
\begin{equation}
\langle \Psi_{\boldsymbol{k}_1, \boldsymbol{k}_2}^{(-)}|
\Psi_{\boldsymbol{k}'_1, \boldsymbol{k}'_2}^{(-)}\rangle
=\delta(\boldsymbol{k}_1-\boldsymbol{k}'_1)
\delta(\boldsymbol{k}_2- \boldsymbol{k}'_2) 
+ \delta(\boldsymbol{k}_1-\boldsymbol{k}'_2)
\delta(\boldsymbol{k}_2- \boldsymbol{k}'_1).
\end{equation} 

In the present work, we first generated the one-electron Coulomb functions $\varphi_{kl}^{(c)}(r)$ 
by employing the routine {\tt COULFG} of Barnett~\cite{Barnett96}, on the same grid used for the
$B$-splines.  This choice guarantees a self-consistent grid representation of the time-evolved
wave\-packet and the final continuum states. We then expressed the function in terms of our $B$-spline basis
in order to use the standard integration schemes in that basis.

It is important to remember that there are two indistinguishable electrons in
the final channel, and hence the energy sharings described by $\alpha$ and $\pi/2-\alpha$
represent the same observable event. Therefore, we either need to
consider $0\leq \alpha \leq \pi/4$ or $\pi/4 \leq \alpha \leq
\pi/2$ to avoid double counting. The TDCS
with respect to the energy of one electron is then given by
\begin{eqnarray}
\frac{{\rm d}^{3}\sigma}{{\rm d}E_1{\rm d}\hat{\boldsymbol{k}}
_1{\rm d}\hat{\boldsymbol{k}}_2} &=&
\frac{1}{k_1k_2\cos^2\alpha}
\bigg(\frac{\omega}{I_0}\bigg)^2\frac{1}{T_{\mathrm{eff}}^{(2)}}\int{\rm d}
k'_1{\rm d} k'_2 
k_1'\delta(k'_2-k'_1\tan\alpha\big) \nonumber \\
& & ~~~~~\times \bigg|
\sum_{L=0,2,l_1l_2}\hspace{-3mm}\chi_{l_1l_2}(k'_1,k'_2)
{\cal Y}_{l_1l_2}^{LM}(\hat{\boldsymbol{k}}
_1, \hat{\boldsymbol{k}}_2)
{\cal F}_{l_1l_2}^{L}(k'_1,k'_2) \bigg|^2.
\end{eqnarray}
Here we have defined the two-electron projection 
\begin{equation}
{\cal F}_{l_1l_2}^{L}(k_1,k_2) \equiv  \langle \Psi_{k_1 l_1,k_2 l_2}^{L(-)}|\Psi(t)\rangle,
\label{eq:mom2}
\end{equation}
with the phase factor 
\begin{equation}
\chi_{l_1l_2}(k_1,k_2)=(-{\mbox i})^{l_1+l_2}e^{{\mathrm i}(\sigma_{l_1}(k_1)+\sigma_{l_2}(k_2))},
\end{equation}
and
\begin{equation}
\Psi_{k_1 l_1,k_2 l_2}^{L(-)} \equiv {\cal A} \left\{\varphi^{(c)}_{k_1
l_1}(r_1)\varphi^{(c)}_{k_2 l_2}(r_2) 
               {\cal Y}_{l_1 l_2}^{L}(\hat{\boldsymbol{k}}_1,\hat{\boldsymbol{k}}_2)\right\}
\end{equation}
as the properly anti\-symmetrized product of two one-electron radial Coulomb functions and
coupled spherical harmonics  ${\cal Y}_{l_1 l_2}^{L}$ for the angular part. [Since $M=0$ for linearly
polarized radiation, we omit it to simplify the notation.]

By collecting all ionization events, the total cross
section for two-photon double ionization is obtained as
\begin{equation}
\fl
\sigma_{\rm NS} = \int{\rm d}\alpha\int{\rm 
d}\hat{\boldsymbol{k}}_1{\rm d}\hat{\boldsymbol{k}}_2
\frac{{\rm d}^3\sigma}{{\rm d}\alpha{\rm 
d}\hat{\boldsymbol{k}}_1{\rm d}\hat{\boldsymbol{k}}_2} 
=
\bigg(\frac{\omega}{I_0}\bigg)^2\frac{1}{T_{\mathrm{eff}}^{(2)}}
 \int{\rm d} k_1{\rm d} k_2
\frac{{\rm d}^2{\cal P}(k_1,k_2)}{{\rm d}k_1{\rm d}k_2},
\end{equation}
where we have introduced the momentum distribution
\begin{equation}
\frac{{\rm d}^2{\cal P}(k_1,k_2)}{{\rm d}k_1{\rm d}k_2}
=\sum_{L=0,2}\sum_{l_1l_2}
\big|{\cal F}_{l_1l_2}^L(k_1,k_2)|^2.
\end{equation}
Similarly, the momentum distribution is given by
\begin{equation}
\frac{{\rm d}^2{\cal E}(E_1,E_2)}{{\rm d}E_1{\rm d}E_2} =
\frac{1}{k_1k_2}\frac{{\rm d}^2{\cal P}(k_1,k_2)}
{{\rm d}k_1{\rm d}k_2}.
\end{equation}

Below we will also show results obtained for two linearly polarized XUX laser pulses,
for which the electric field is given by
\begin{equation}
\mbox{\boldmath$E$}(t)=\big[E_1f_1(t)\sin(\omega_1 t)
+E_2f_2(t-T_d)\sin(\omega_2(t-T_d)) \big] \mbox{\boldmath$e$}_z.
\end{equation}
Here $\omega_1$ and  $\omega_2$ are the two central frequencies while 
$T_d$ is the time delay between the two pulses. The total time duration
can be written as
$\tau = \max\{\tau_1,\tau_2+T_d\}$.

Note that this time duration depends on the durations of pulse~1, 
pulse~2, and the time delay between them. Hence, the definition of
an ``effective interaction time'' is by no means straightforward and 
even a ``generalized cross section'' (which should be independent of these parameters)
cannot be defined in the standard sense. Momentum distributions, however, 
can be obtained by calculating and then squaring the magnitude of 
the function~(\ref{eq:mom2}).  Energy distributions for the He case were recently reported by
Foumouo \etal\cite{Foumouo2008}.

\section{RESULTS AND DISCUSSION}

As a first test of our newly developed approach, we applied it to the single-color
two-photon double ionization of helium, for which a variety of benchmark data exist 
for comparison (see, for example, refs.~\cite{Feist2008,Guan2008a} and references therein).
In this particular test, we confined the system to a spatial box of $r_{\rm max}=60\,$a.u.
We used 81 $B$-splines of order~8 to span this configuration space, with 
the knot sequence chosen in such a way that the spatial variation of the
wave\-function close to the nucleus (determined by the nuclear charge~$Z$) and far away from the
nucleus (determined by the highest energy of a free electron) could be represented accurately.

Having obtained the initial state of the system by diagonalizing the field-free Hamiltonian
of the $\rm ^1S^e$ symmetry and taking the eigenvector to the lowest eigenvalue ($-2.90330\,$a.u.~in our
particular case), the time propagation in the 
laser field is accomplished through the Arnoldi-Lanczos algorithm. Compared to other
time-propagation approaches, such as leapfrog or a split-operator approach~\cite{Hu2005}, 
the present scheme allows us to take relatively large steps in time.  Specifically,
using only 100 steps per optical cycle (o.c.) 
is sufficient to achieve converged solutions of the TDSE for the cases presented in this paper. 

The results presented below were obtained with a sine-squared pulse with a  peak intensity of
$5\times 10^{14}\,$W/m$^2$ and a time duration of 10~optical cycles. For this case, we first checked 
the total cross sections for photon energies of $42$ and $50\,$eV, respectively.  These values were 
chosen to ensure that the box size and the length of the laser pulse were suitable to properly define 
a generalized cross section for two-photon double ionization.  The present results
of $\rm 4.02\,\times 10^{-53}\,cm^4 s$ and $\rm 2.20\,\times 10^{-52}\,cm^4 s$ agree in 
a satisfactory way with the corresponding values of $\rm 3.76\,\times 10^{-53}\,cm^4 s$ 
and $\rm 1.89\,\times 10^{-52}\,cm^4 s$ obtained in an entirely different finite-element
discrete-variable representation (FE-DVR)~\cite{Guan2008a}.

\subsection{Triple-differential cross sections for one-color two-photon double ionization}
\begin{figure*}[htb]
\centering
~~~~~\epsfig{file=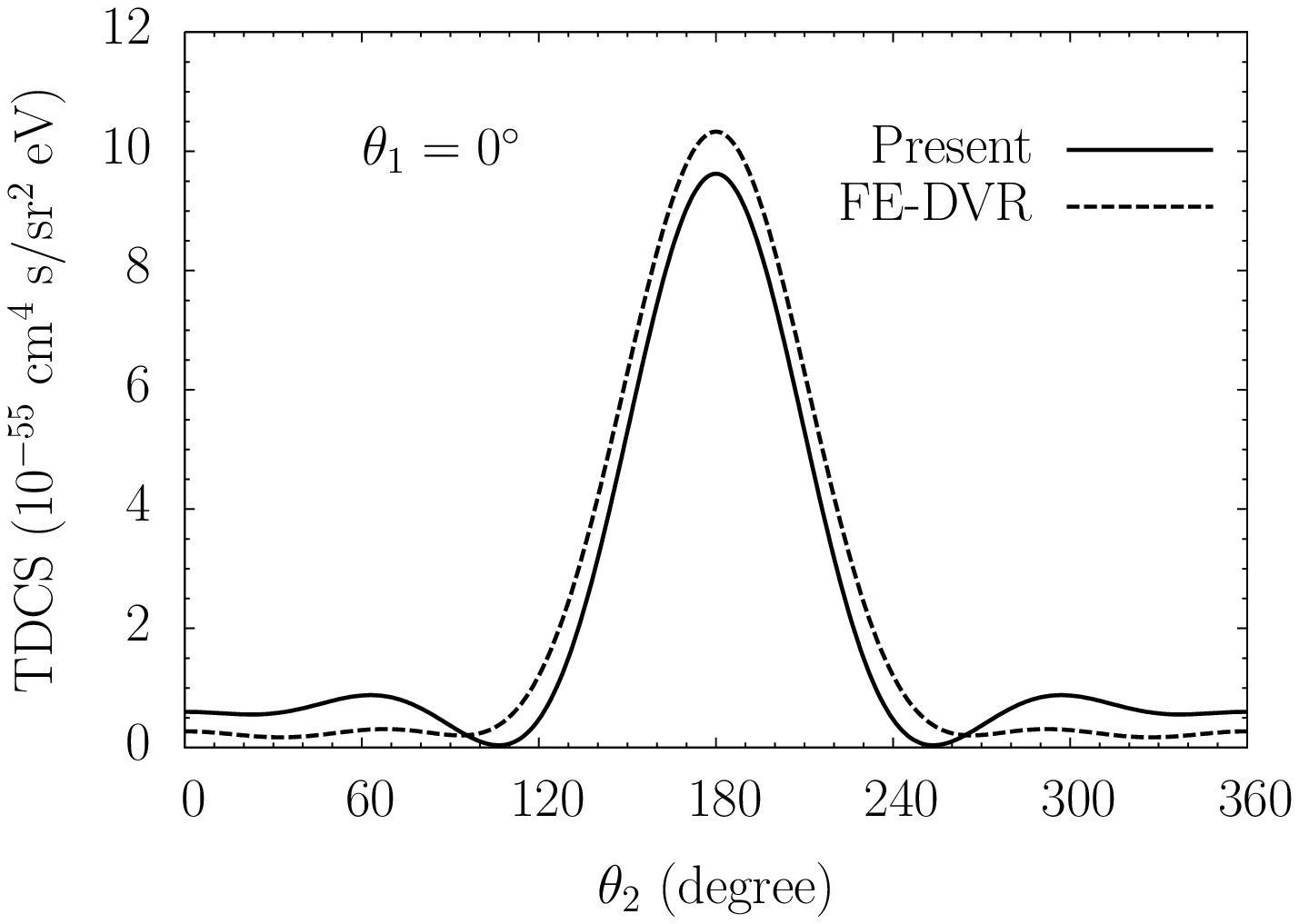,width=6.5cm}
\epsfig{file=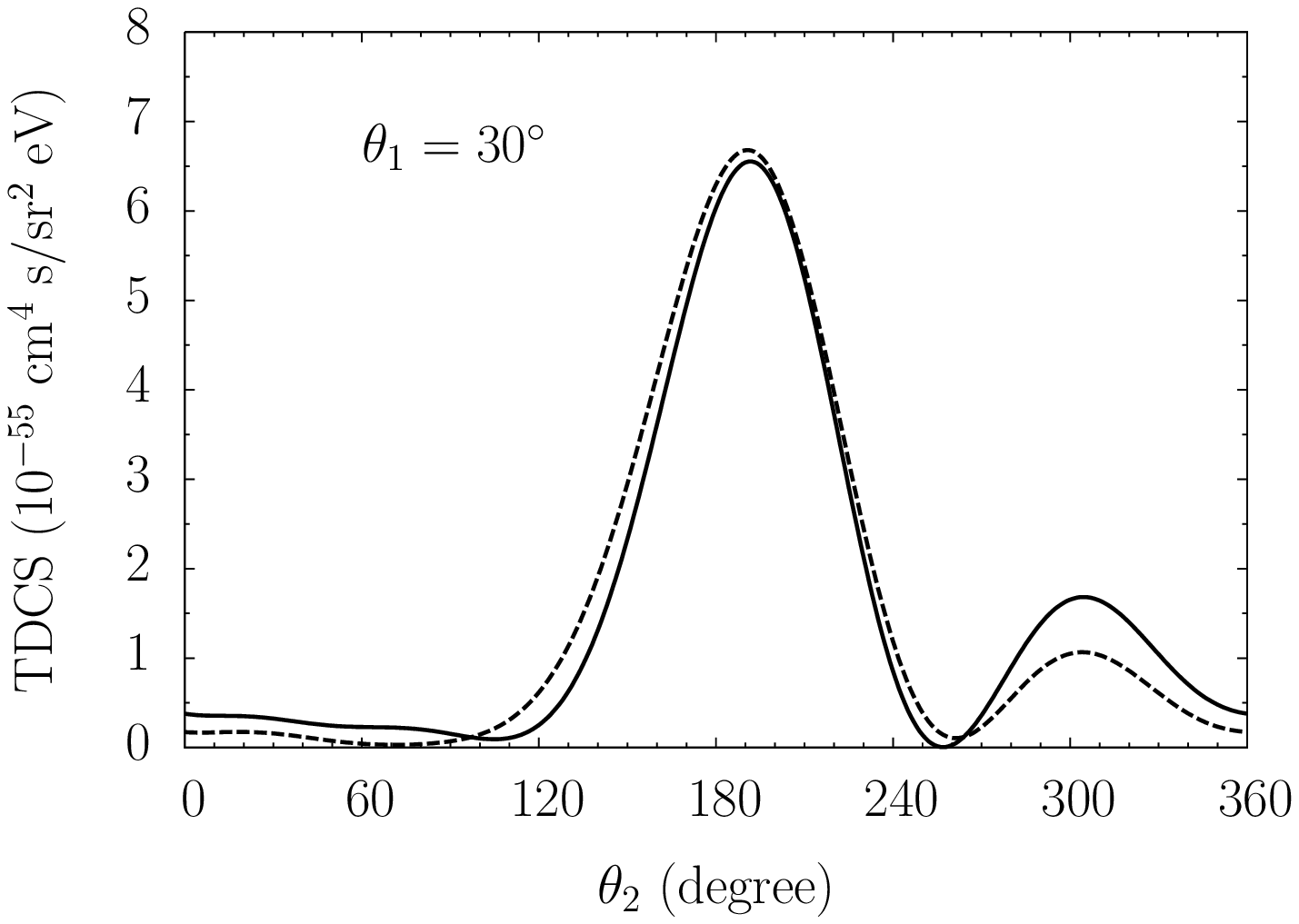,width=6.5cm} \\
~~~~~\epsfig{file=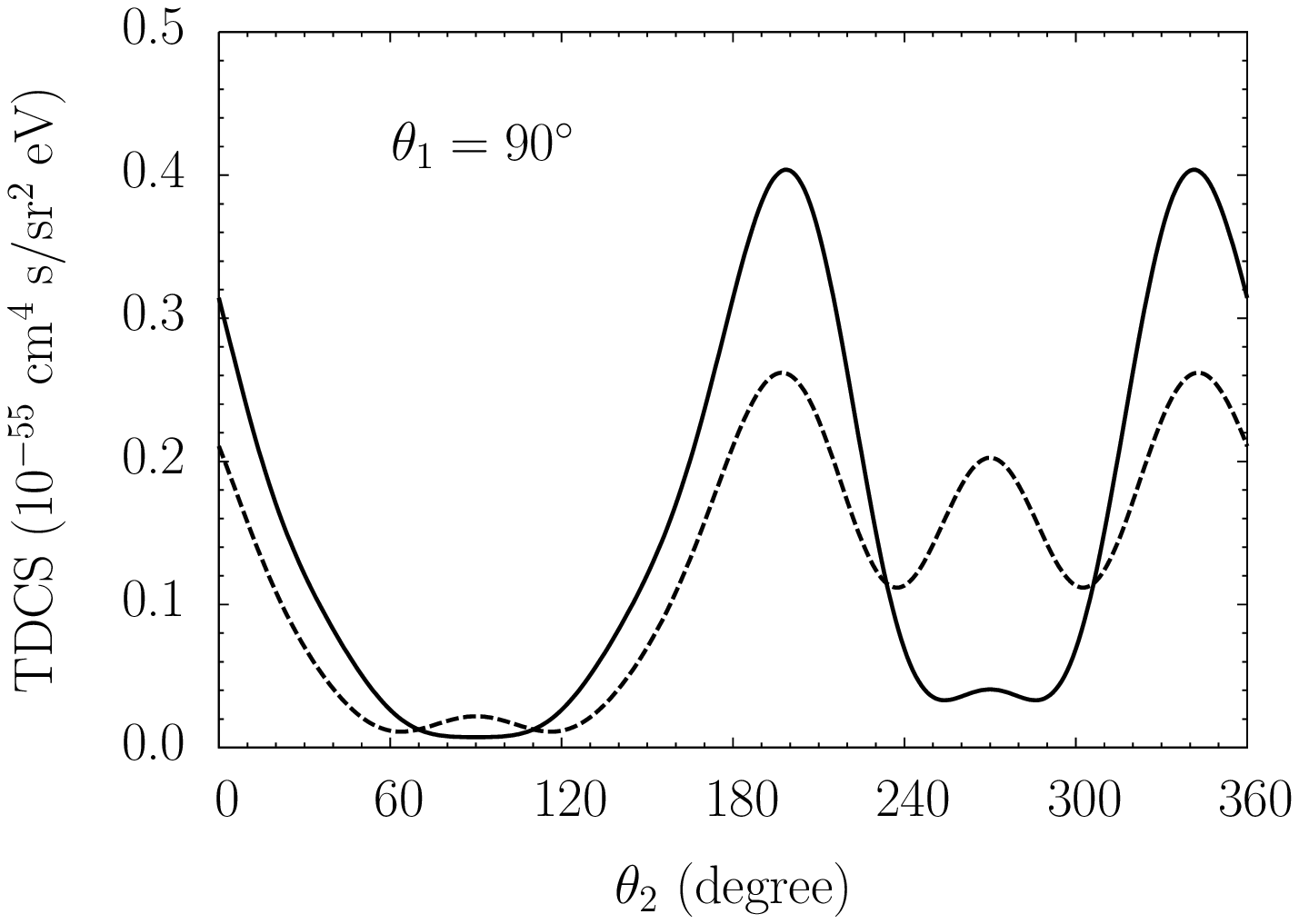,width=6.5cm}
\epsfig{file=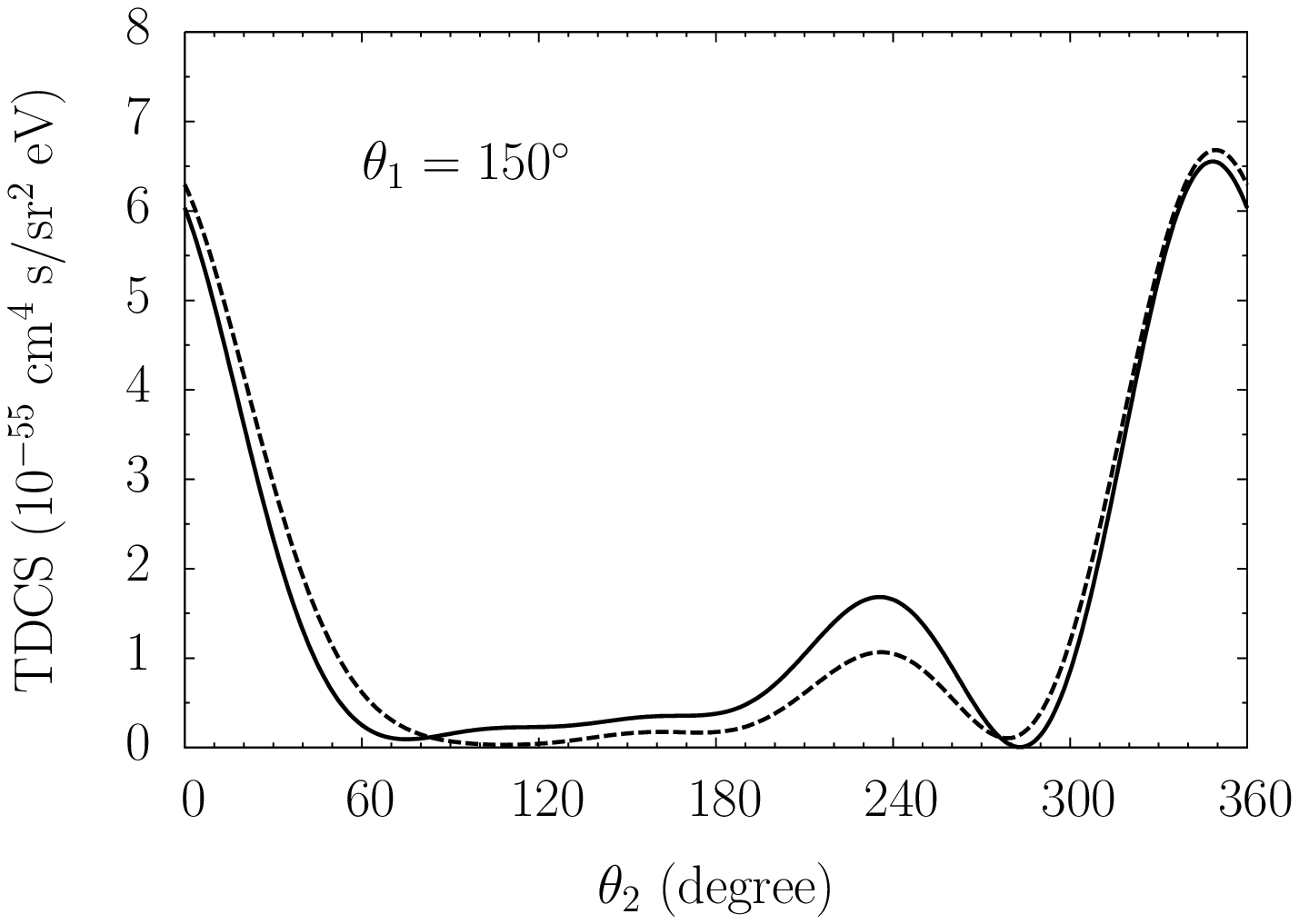,width=6.5cm} 
\caption{The coplanar triple-differential cross section for
two-photon double ionization of helium in a 10-cycle sine-squared laser pulse of 
central photon energy $42\,$eV and peak intensity $5\times 10^{14}\,$W/cm$^2$ for 
equal energy sharing
($E_1=E_2=2.5\,$eV) of the two outgoing electrons.  The angle listed in the
figure is the angle between the laser polarization vector and one of the two escaping electrons, while
the emission angle of the second electron varies. The present TDBSR results are compared with 
those from our previous FE-DVR calculation~\cite{Guan2008a}.}
\label{fig:tdcs-coplanar}
\end{figure*}
Figure~\ref{fig:tdcs-coplanar} displays the triple-differential cross section for
two-photon double ionization of helium in a 10-cycle sine-squared laser pulse of 
central photon energy $42\,$eV and peak intensity \hbox{$5\times 10^{14}\,$W/cm$^2$} for 
equal energy sharing
\hbox{($E_1=E_2=2.5\,$eV)} of the two outgoing electrons.  These results are for the coplanar 
geometry, where the electric field vector of the linearly polarized laser field and the
momentum vectors of the two escaping electrons all lie in the same plane. 

Given the very different implementations of the current TDBSR method and the
FE-DVR approach, we 
see satisfactory agreement between the present results and those obtained earlier. Note that these results
were obtained with  $(L_{\rm max},l_{1,\rm max},l_{2,\rm max}) = (3,3,3)$.  We are aware of potential
convergence problems with these parameters, especially for kinematic situations where the cross section
is small (e.g., at $\theta_1 = 90^\circ$).  However, since the principal motivation for performing these particular 
calculations was to test the method and the accompanying computer code, we consider the test successful 
and hence move on to the two-color-case in the next sub\-section.

\subsection{Momentum and energy distributions for two-color two-photon double
ionization}
We now consider the process of double photo\-ionization by absorption
of two photons at different central photon energies. In other words, the target
helium atom is exposed to the irradiation by two laser pulses, of potentially different frequencies and
with a controllable time delay. This
allows us to study the mechanism of the breakup problem by applying pulses
of various durations for each pulse and also modifying the time delay between them. 
The following two-color laser parameters were considered in this work:
pulse~1 has a central photon energy of $35.3\,$eV, a time duration of $12\,$o.c., and a
peak intensity of $1\times 10^{14}\,$W/cm$^2$, while the corresponding parameters
for pulse~2 are $57.1\,$eV, $14\,$o.c., and $1\times 10^{13}\,$W/cm$^2$, respectively.
A similar two-color laser pulse was studied in the recent work by Foumouo \etal\cite{Foumouo2008}.  

\begin{figure}[t]
\centering
\epsfig{file=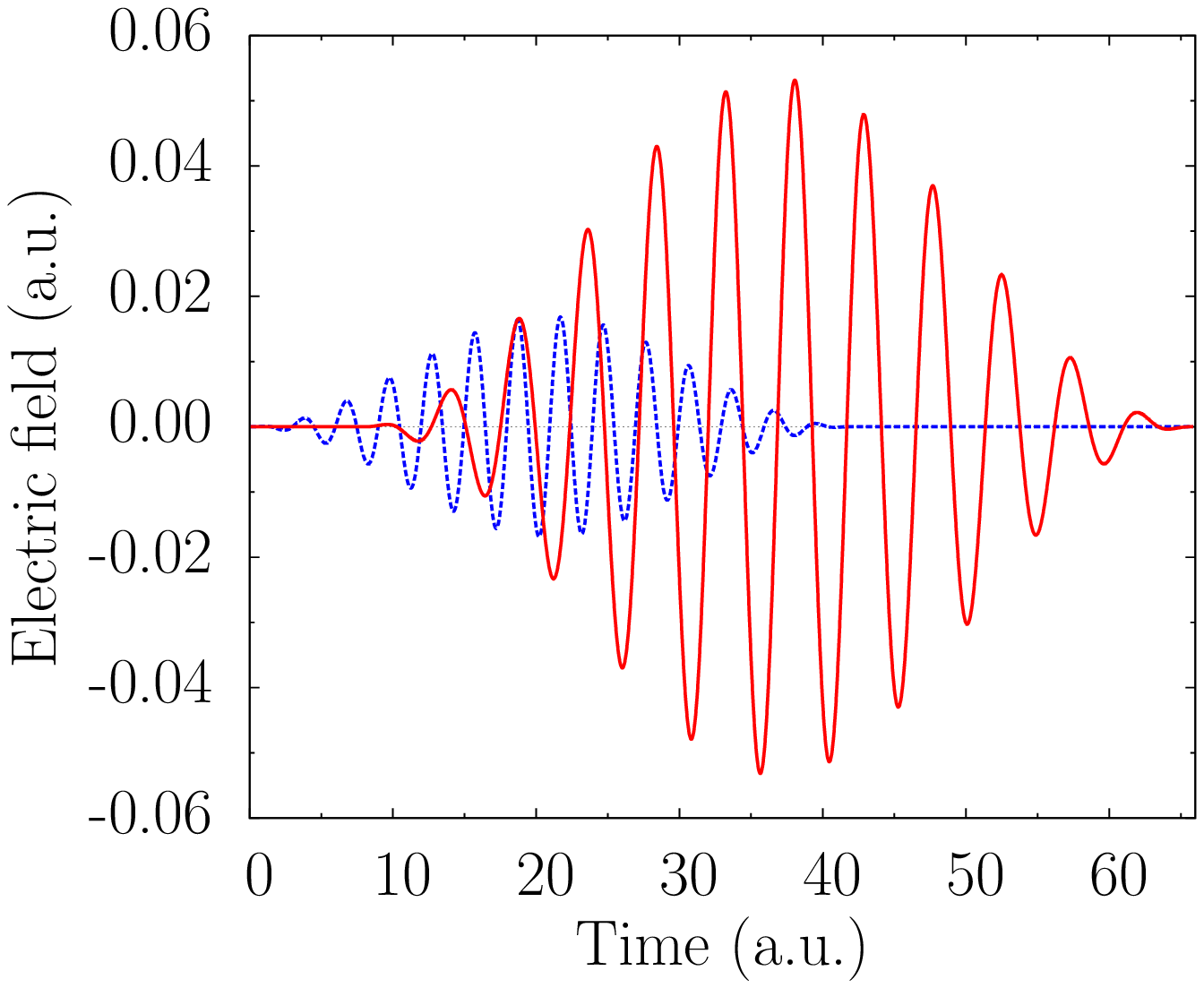,width=5.cm}
\epsfig{file=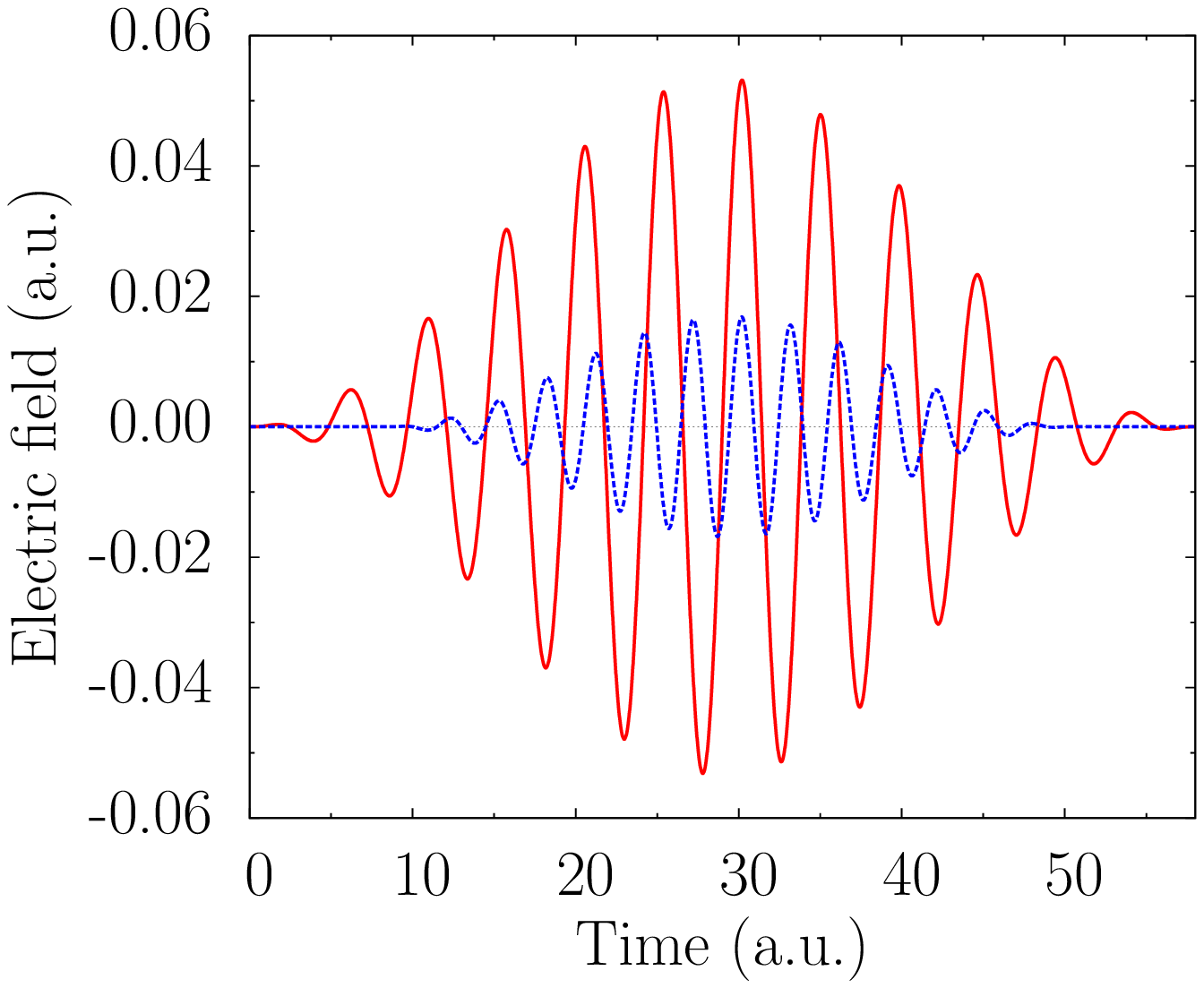,width=5.cm} 
\epsfig{file=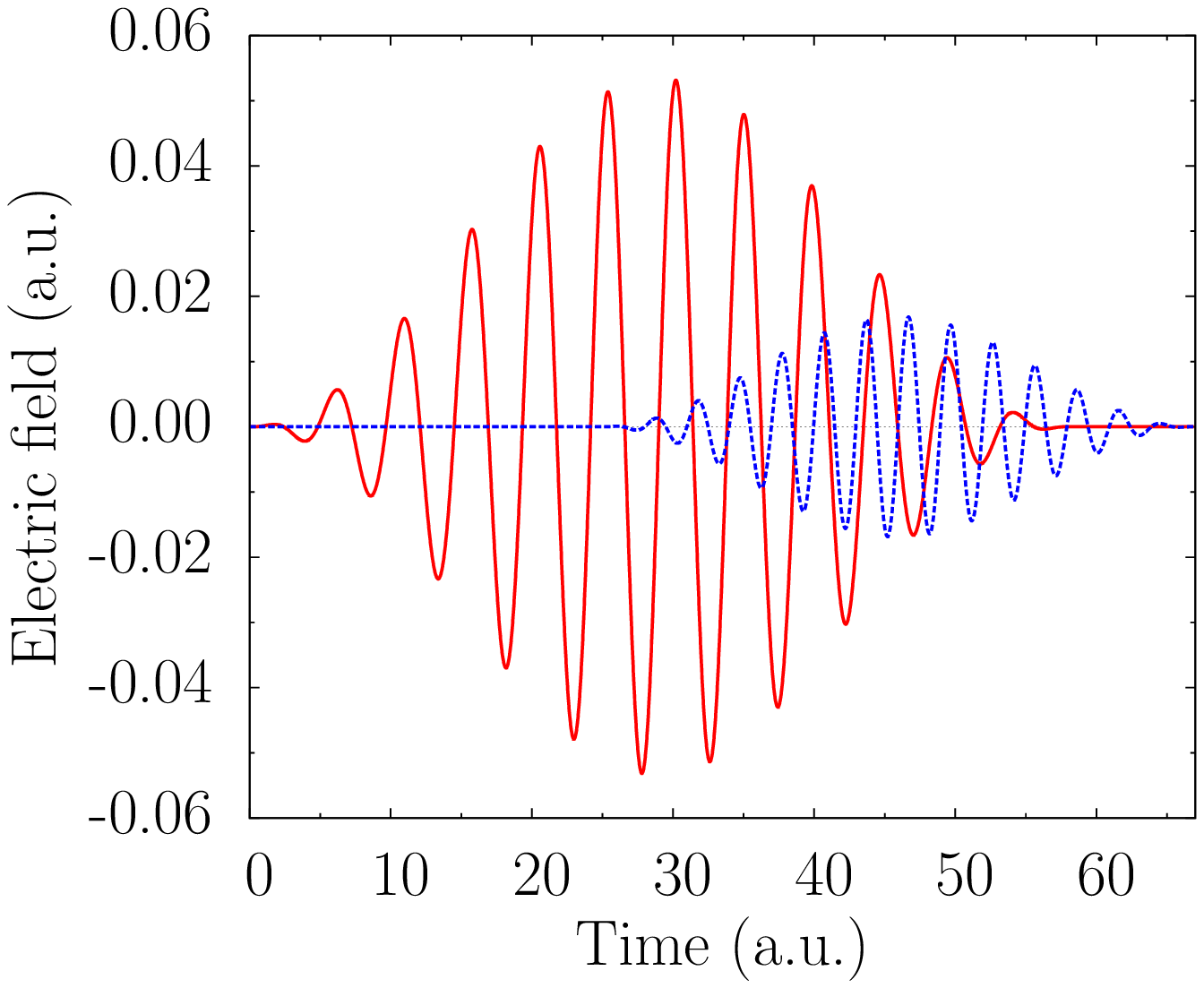,width=5.cm}
\caption{ Electric field of the two-color laser pulse. The laser parameters
are: $\omega_1=35.3\,$eV, $\tau_1=12\,$o.c.~at a peak intensity of
$10^{14}\,$W/cm$^2$ (solid curve), and $\omega_2=57.1\,$eV, $\tau_2=14\,$o.c.~at
a peak intensity of $10^{13}\,$W/cm$^2$ (dashed curve).
The time delays between the maxima of the two pulses are, from left to right, $-16.5$, $0.0$,
and 16.5$\,$a.u.}
\label{fig:ele-twoc}
\end{figure}

In the present work, we are interested in the mechanism for the ejection of two electrons
when the time delay between the two pulses is varied. Figure \ref{fig:ele-twoc} displays
the time-dependent electric fields of the two-color pulses for time delays
of $-16.5\,$a.u. (i.e., pulse~2 comes first), no delay (both pulses come simultaneously, and 
a delay of $+16.5\,$a.u.~(pulse~1 comes first).  
Note that $+16.5\,$a.u correspond to approximately 400 atto\-seconds. 

\begin{figure}[t]
\centering
\epsfig{file=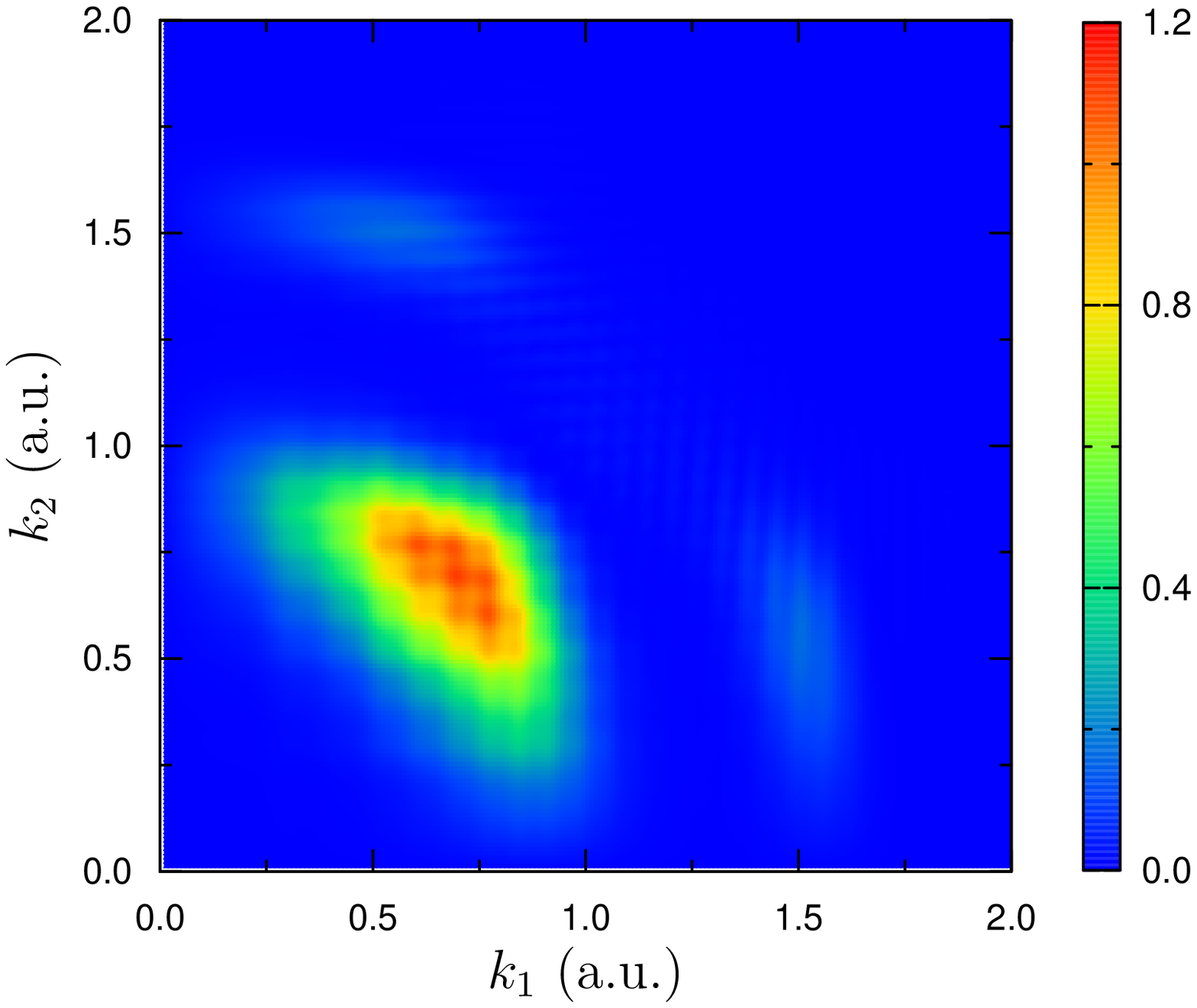,angle=-90,width=6.2cm}
\epsfig{file=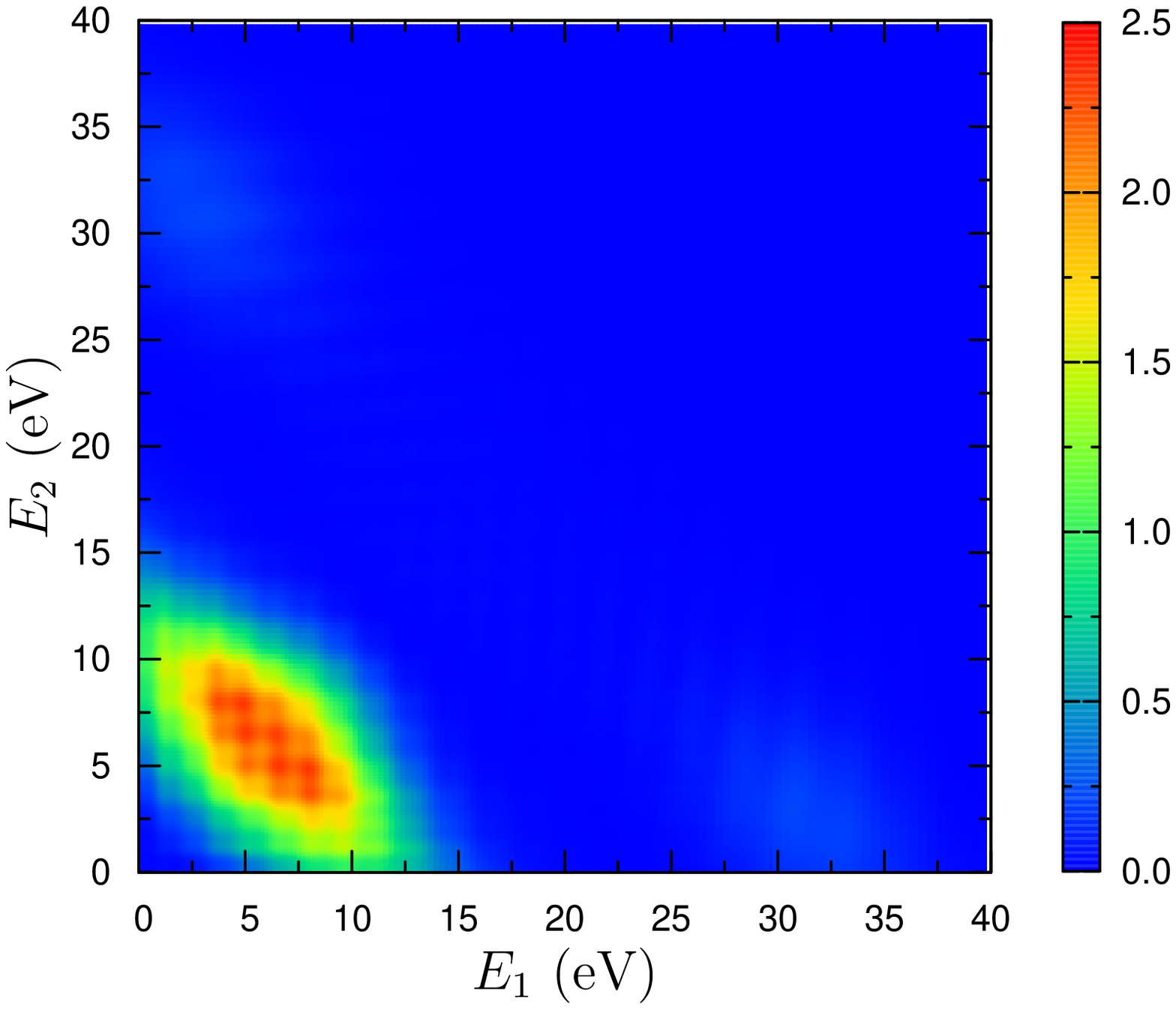,angle=-90,width=6.2cm}\\
\epsfig{file=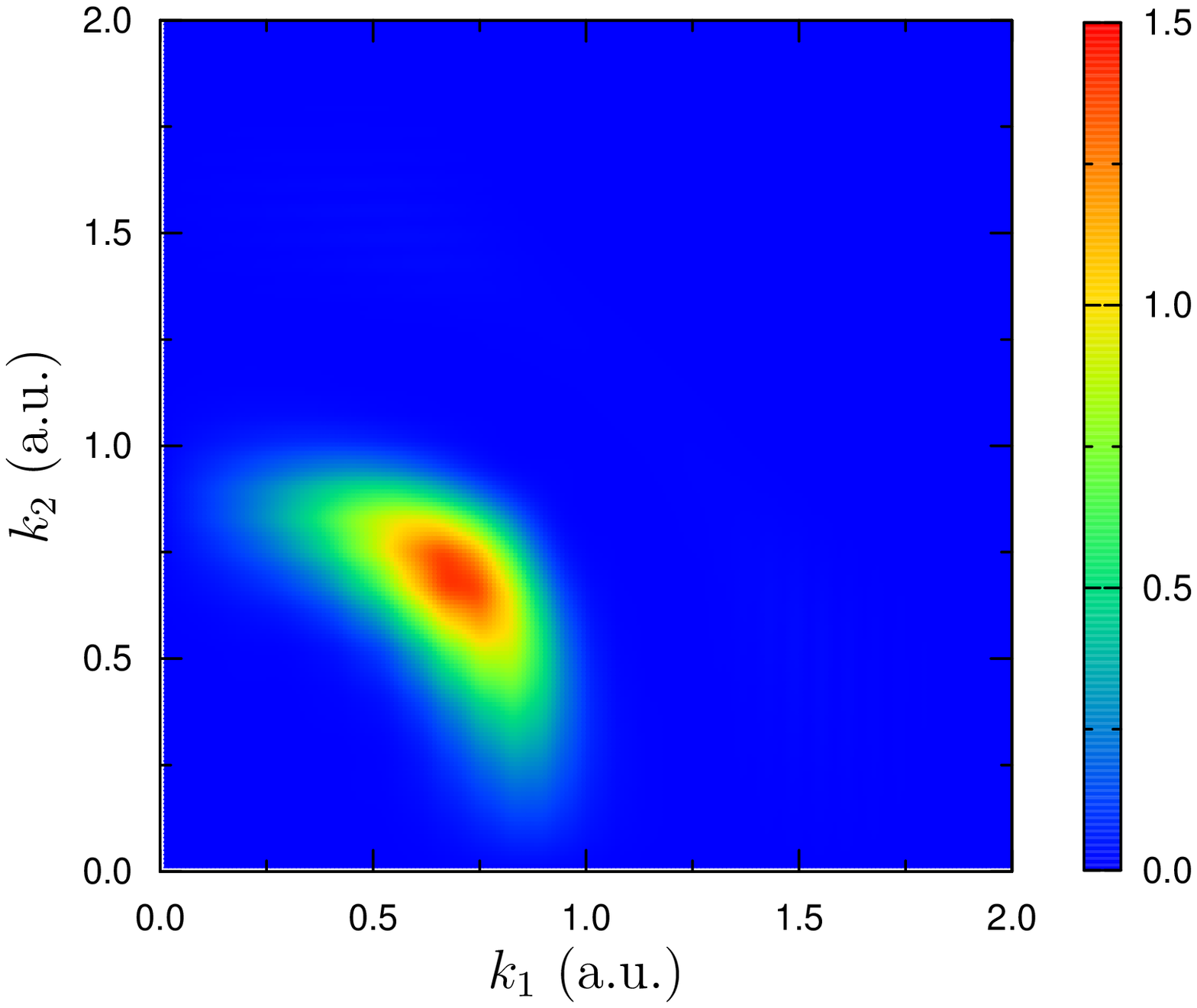,angle=-90,width=6.2cm}
\epsfig{file=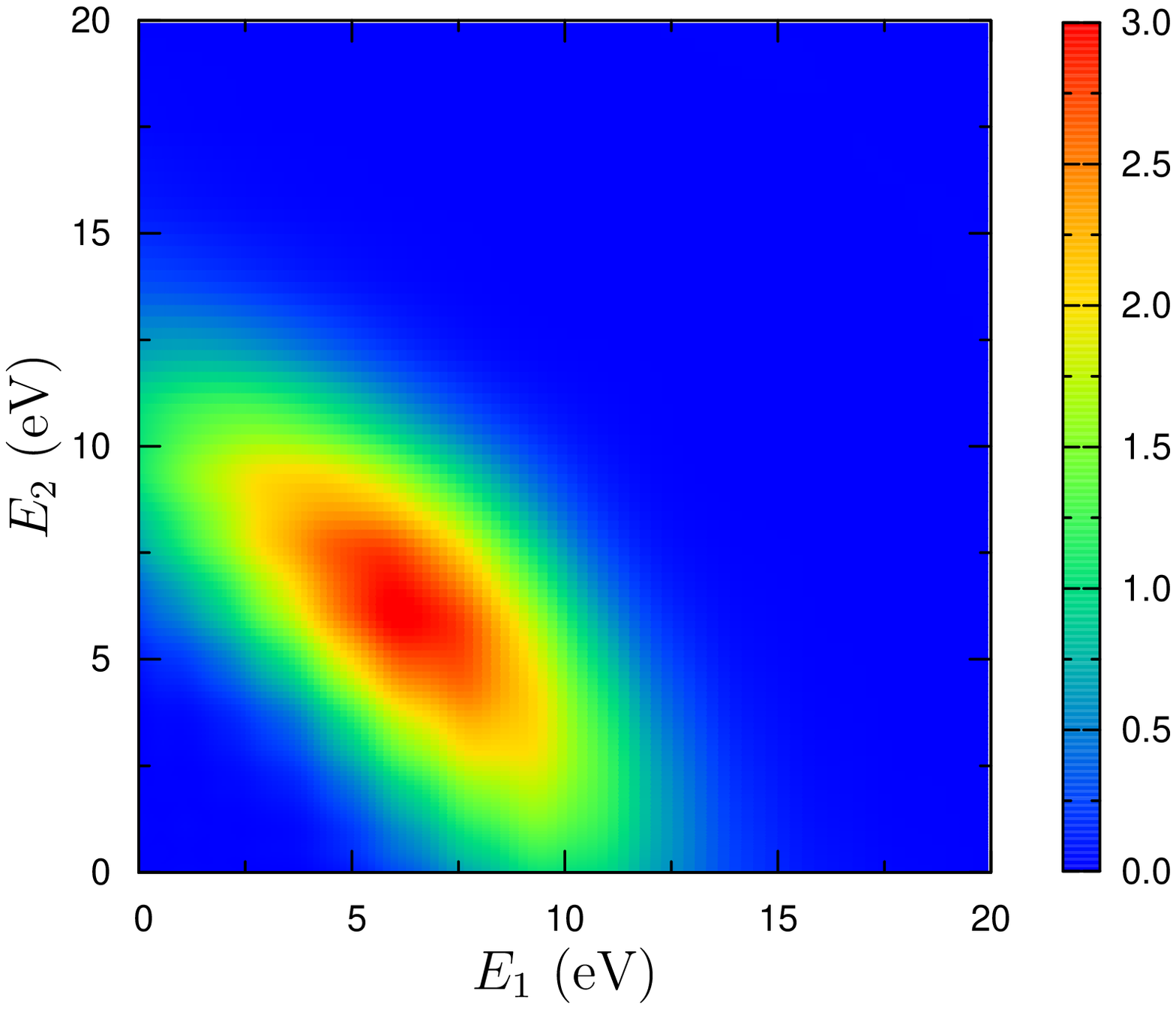,angle=-90,width=6.2cm}\\
\epsfig{file=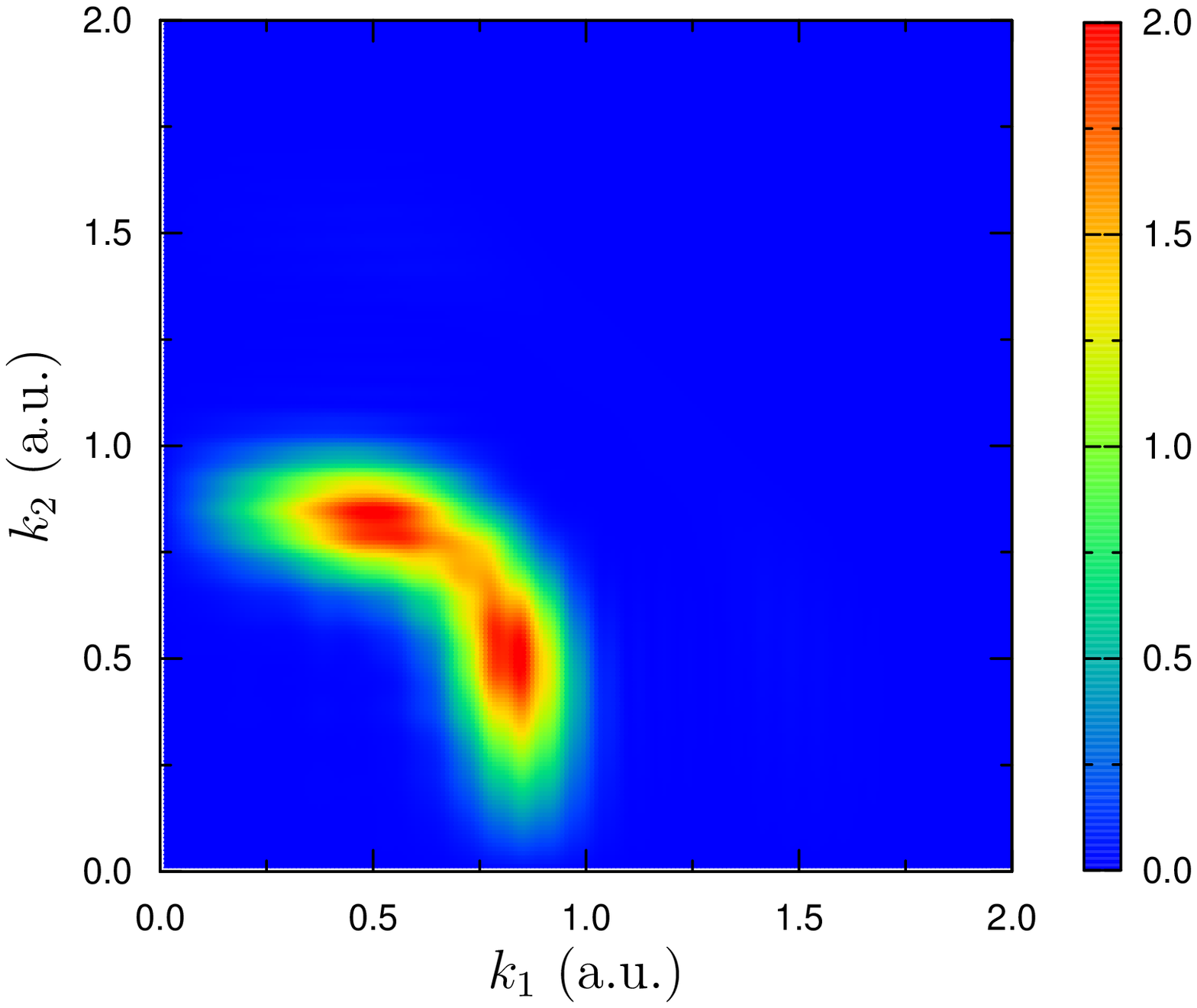,angle=-90,width=6.2cm}
\epsfig{file=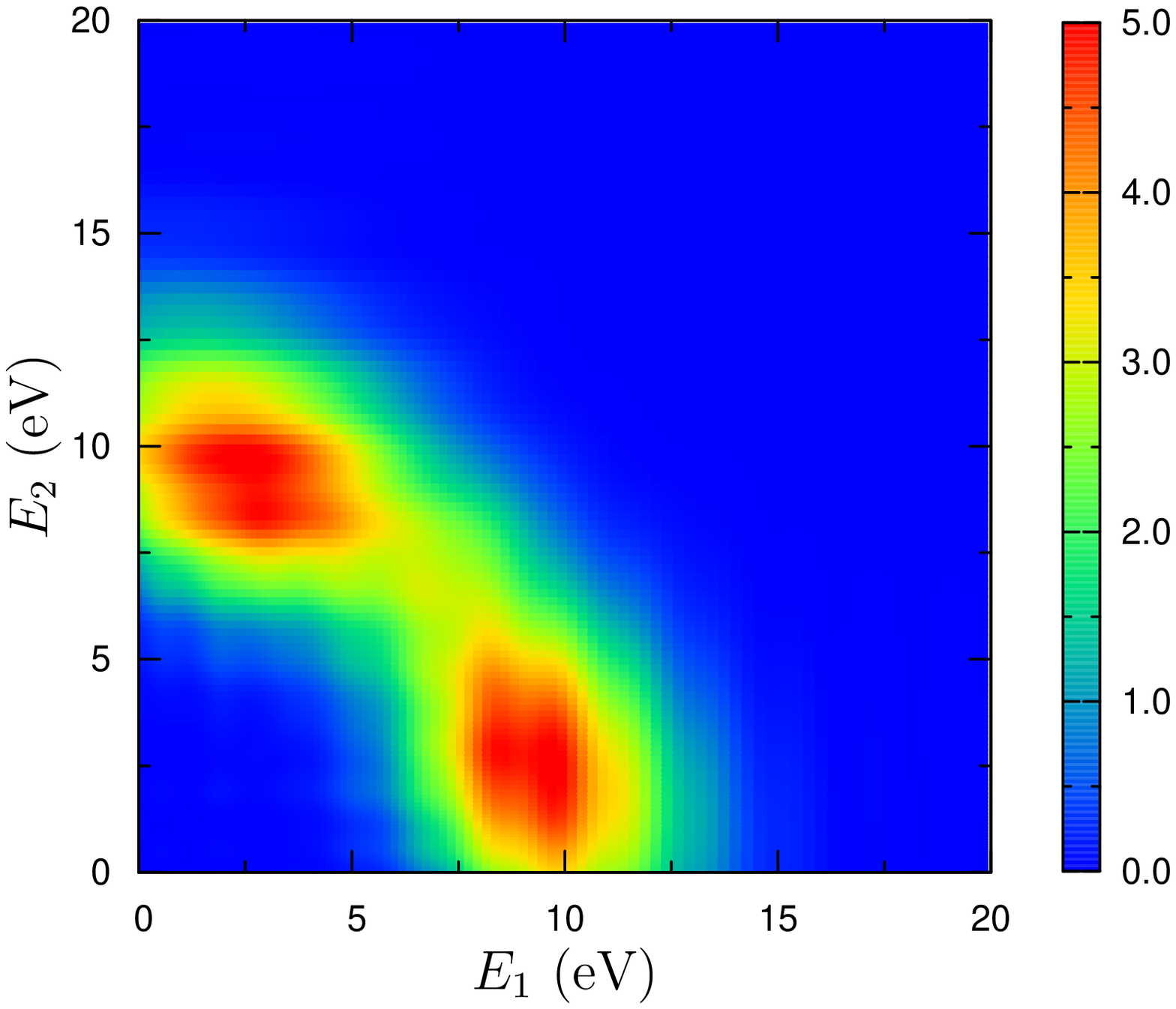,angle=-90,width=6.2cm}
\caption{Momentum (left column) and energy (right column) distributions of
the two escaping electrons in a two-color laser pulse. The laser parameters are
the same as in figure~\ref{fig:ele-twoc}.
The color bars for the first row (negative time delay) correspond to
units of $10^{-5}\,$a.u., while the other two rows are given in units of $10^{-4}\,$a.u.
The time delays
between the two laser pulses, from top to bottom, are $-16.5$, $0.0$, and $+16.5\,$a.u., respectively.
}
\label{fig:twoc-ene}
\end{figure}

Figure \ref{fig:twoc-ene} depicts the momentum and energy distributions of the two
escaping electrons for the above two-color laser pulses. 
For the negative time delay with pulse~2 arriving first, we
observe a strong peak around $E_1=E_2\simeq 6.7\,$eV, as well as two 
weaker peaks around $(E_1,E_2)\simeq (2.7, 32)\,$eV and $(E_1,E_2)\simeq (32,2.7)\,$eV, respectively. 
This suggest that the target electrons predominantly absorb the two
photons in the highly correlated way similar to the {\it nonsequential\/} double ionization mechanism,
even though the photon energies are different. Recent calculations by Feist \etal\cite{Feist2009}, using a single-color laser,
also demonstrate that the conventional scenario of ``direct'' vs ``sequential'' two-photon double 
ionization breaks down for atto\-second XUV pulses, even when the sequential process is energetically allowed.

In the above two-color case, the two ejected
electrons share the excess energy of $E_{\rm exc}= \omega_1+\omega_2 - I^{2+}\simeq 13.5\,$eV
almost equally. However, there is apparently another
ionization channel open for the present laser parameters. Since the target
electrons interact with pulse~2 of energy 57.1$\,$eV first, 
the first electron is ejected with a kinetic energy of about $32\ (\simeq 57.1-24.6)\,$eV before
the second electron is ionized  with an energy of $2.7\ (\simeq 57.1-54.4)\,$eV from the ground state 
of the residual He$^+$(1s) ion. This
channel is characteristic of sequential ionization 
{\it by two photons of the same high frequency\/} and very 
different from the direct channel. In fact, in this case there is no electron left for the second pulse
to interact with.

To improve the visibility of the remaining figures, we will 
show the energy distributions only in the energy region below $20\,$eV
for the cases of zero and positive time delays. 
When there is no time delay between the two maxima of the electric fields, the
principal structure observed is still around $E_1=E_2\simeq 6.7\,$eV in the
energy distribution. The mechanism in this case is thus similar to what we observed for
the negative time delay. 

On the other hand, when the time delay is increased further, for
example up to $+16.5\,$a.u., the two electrons are ejected a
very different way. In this case, the main structure revealed from figure~\ref{fig:twoc-ene} 
exhibits two peaks at $(E_1,E_2)\simeq (10.0,2.5)\,$eV and $(2.5,10.0)\,$eV. This is 
due to the fact that pulse~1 comes first and hence, after
absorption of one photon of frequency~($\omega_1$) one electron with the kinetic
energy of $10.7~(=35.3-24.6)\,$eV is ejected. The time delay of
+16.5$\,$a.u.~is sufficiently long for the residual He$^+$ ion to relax into its
ground state after the single ionization. When this process is followed by the 
absorption of the second photon of $57.1\,$eV ($\omega_2$), the other
electron carries away the excess energy of $2.7=(57.1-54.4)\,$eV. This suggests that
the two electrons are basically removed through a sequential process. 

\begin{figure}[t]
\centering
\epsfig{file=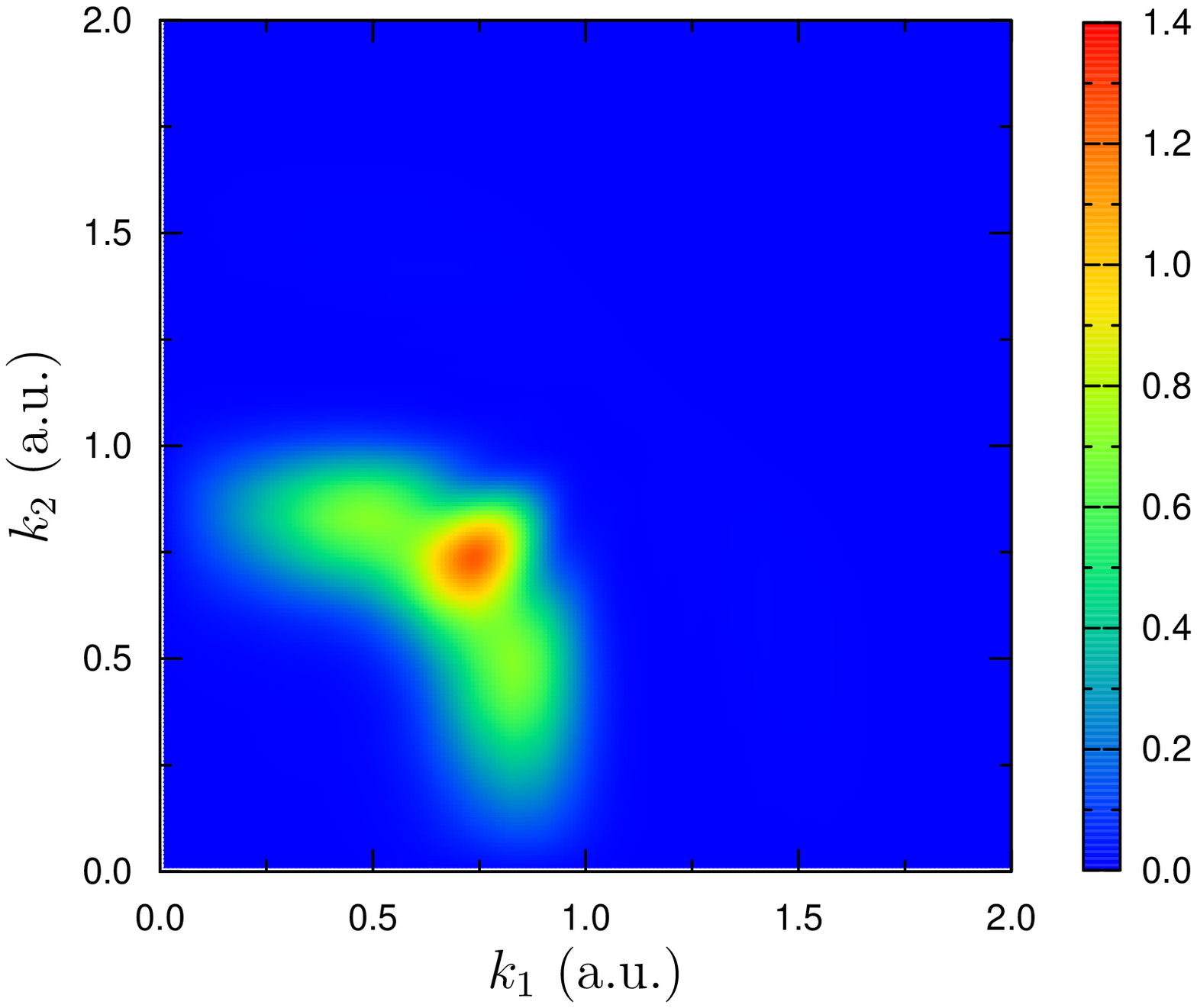,angle=-90,width=6.2cm}
\epsfig{file=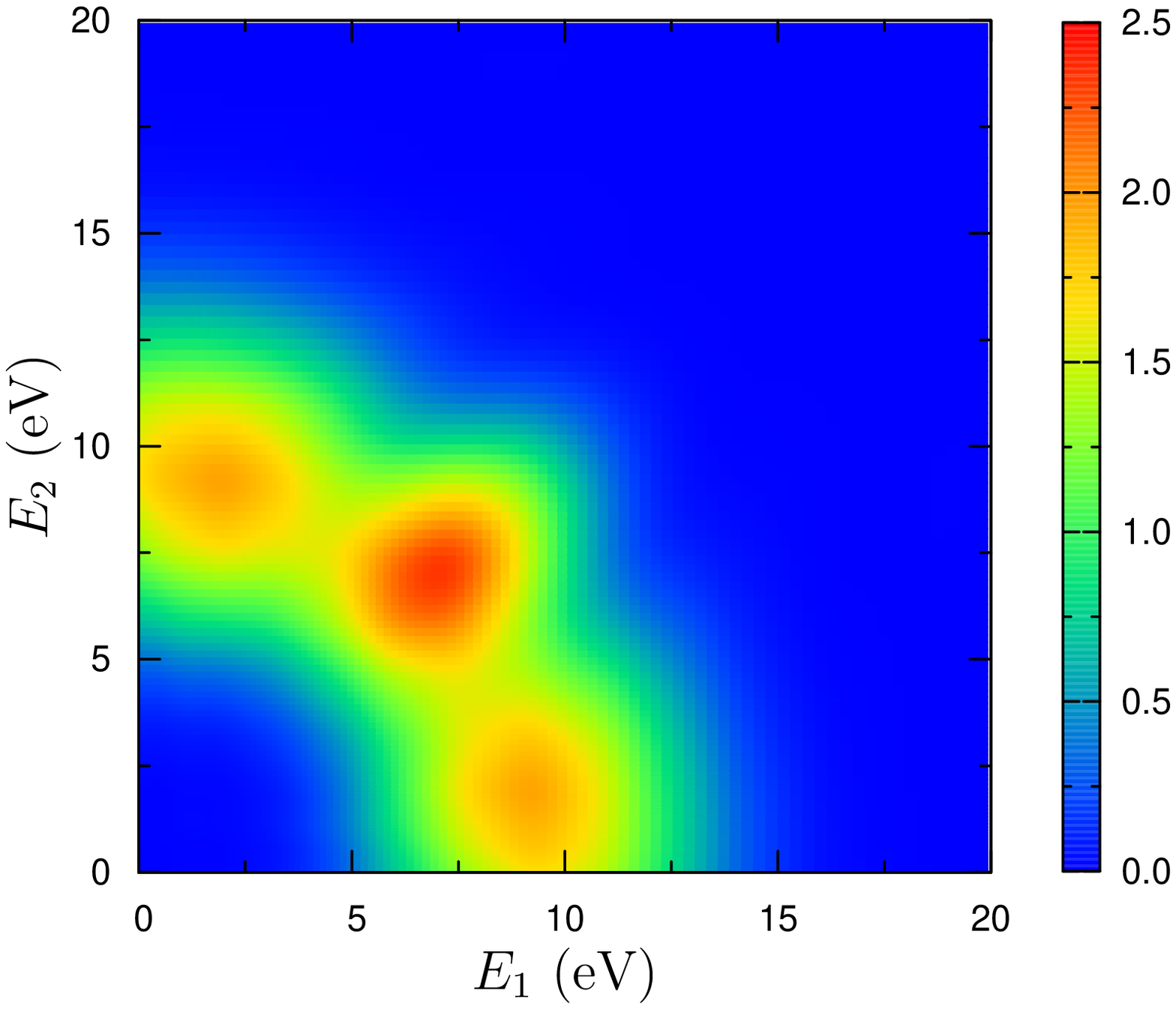,angle=-90,width=6.2cm}\\
\caption{Momentum (left) and energy (right) distributions of
the two escaping electrons in a two-color laser pulse. The wavelengths and laser intensities are
the same as in figure~\ref{fig:ele-twoc}, but in this case the pulse\-lengths are only 10 optical
cycles with a delay of 15~a.u.
The color bars are given in units of $10^{-4}\,$a.u.
}
\label{fig:twoc-ene2}
\end{figure}

We then explored the parameter space in more detail by varying both the pulselength
and the delay. An even more striking example of the competing processes is exhibited 
in figure~\ref{fig:twoc-ene2}.  For the same wavelengths and intensities as before, but
with a relatively small change in the pulse\-lengths to 10 optical
cycles and a delay of 15~a.u. ($\approx 360\,$atto\-seconds), sequential and non-sequential
absorption of the two photons play an almost equally important role in the double
ionization process.  
 
From the discussion and examples given above, it is clear that the time delay plays a decisive
role in determining how the two electrons are ejected by two-color XUV laser pulses.
Depending on the details of time delay, the electrons can be ejected in ways either
similar to the sequential or the non-sequential process. Our findings qualitatively agree with 
those of Foumouo \etal~\cite{Foumouo2008} for the two-color problem and Feist \etal\cite{Feist2009} 
in the single-color problem. They serve as an independent confirmation of
their predictions, and also provide us confidence in our newly developed computer code.

\section{CONCLUSIONS}

We have presented a {\it general\/} method to calculate two-photon double ionization of atoms using a 
$B$-spline $R$-matrix approach in connection with an efficient Arnoldi-Lanczos time
propagation scheme.  Test calculations for helium revealed good agreement with previous benchmark 
results obtained with different and entirely independent methods.  The method was then applied to
two-photon double ionization of helium using two time-delayed XUV laser pulses.  The latter results  
confirm that atto\-second spectroscopy will provide a ``microscope'' to examine and also control
the way electrons interact in atomic and molecular targets.

We are currently in the process of generating and then transforming the corresponding matrices for
the two-photon double ionization problem of neon atoms.  
This will allow us to make a direct comparison with the recent experiments of
Moshammer \etal \cite{Mosha2007} carried out at the FLASH facility in Hamburg.
While the additional
complication of a residual core with non-zero angular momentum
is substantial, the current method has been formulated in such a way that 
these calculations are effectively  limited by the available hardware (i.e.~super\-computer facilities)
rather than special-purpose software.  
As long as the field-free and the dipole matrices can first be generated, in any primitive basis, and
then transformed to the eigenbasis, the time propagation essentially becomes a matter of fast 
matrix-vector multiplications, for which highly tuned subroutines are available. 

Consequently, we are confident that we will be able to generate results for complex targets in 
the very near future. In addition, we plan to adapt the method to simple linear molecules that are amenable 
to single-center expansion techniques.

\section*{Acknowledgments}
This work was supported by the United States National Science Foundation
under grants No.~PHY-0757755~(XG and KB) and PHY-0555226 (CJN, OZ, and KB).
We gratefully acknowledge supercomputer resources provided by the U.~S. Department of
Energy through its National Energy Research Scientific Computer Center (NERSC)
and the NSF through Teragrid allocations under TG-PHY090031.

\end{document}